\documentstyle[preprint,aps]{revtex}

\begin{document}
\draft
                     
\title{Tidal Effects Of Passing Planets And Mass Extinctions} 

\author{Daniele Fargion$^{1,2}$ and Arnon Dar$^{3}$}

\address{1.Physics Department, Rome University 1, INFN, Italy\\
         2.Department of Electrical Engineering, Technion, 
           Haifa 32000, Israel\\
         3.Department of Physics and Space Research Institute, 
           Technion, Haifa 32000, Israel}

\maketitle


{\bf Recent observations suggest that many planetary-mass objects may be
present in the outer solar system between the Kuiper belt and the Oort
cloud.  Gravitational perturbations may occasionally bring them into the
inner solar system. Their passage near Earth could have generated gigantic
tidal waves, large volcanic eruptions, sea regressions, large meteoritic
impacts and drastic changes in global climate.  They could have caused the
major biological mass extinctions in the past 600 My as documented in the
geological records.}

Geological records indicate that the exponential diversification of marine
and continental life on Earth in the past 600 My was interrupted by many
mass extinctions (1).  The records also indicate that the major mass
extinctions were correlated in time with large meteoritic impacts,
gigantic volcanic eruptions, sea regressions and drastic changes in global
climate (2-5). Some of these catastrophes coincided in time. The reason for 
that is not clear. Meteoritic
impacts alone, volcanic eruptions alone or sea regressions alone could not
have caused all the major mass extinctions:  A large meteoritic impact was
invoked (6) in order to explain the iridium anomaly and the mass
extinction which killed the dinosaurs and claimed 47\% of existing genera
(1) at the Cretaceous-Tertiary (K/T) boundary 64 $My$ ago.  But, neither
an iridium anomaly, nor a large meteoritic crater have been dated back to
the Permian/Triassic (P/T) mass extinction, $251~My$ ago, which was the
largest known extinction in the history of life (7,8), where global
species extinction ranged between 80\% to 95\%. The gigantic Deccan basalts
flood in India that occurred around the K/T boundary (2,3) and the
gigantic Siberian basalts flood that occurred around the P/T boundary have
ejected approximately $ 2\times 10^6~km^3$ of lava (4). They were more
than a thousand time larger than any other known eruption on Earth, making
it unlikely that the other major mass extinctions, which are of a similar
magnitude, were produced by volcanic eruptions.  However, although there
is no one-to-one correspondence between mass extinctions, large volcanic
eruptions, large meteoritic impacts, and global environmental and climatic
changes, there are clear time correlations between them whose origin is
not clear. Here we suggest that tidal effects of planetary-mass objects
which pass near Earth could have caused all of these and can explain the
complex geological records of the major biological mass extinctions. In
fact, frictional heating of planetary interiors by gravitational tidal
forces can lead to strong volcanic activity, as seen, for instance, on
Jupiter's moon Io, the most volcanically active object known in the solar
system (9,10). But, the moon, the sun and the known planets are too far
away to induce volcanic eruptions on Earth. However, recent observations
(11,12) and theoretical considerations suggest that thousands of
planet-mass objects may be present in a distant ring in the solar system
between the Kuiper belt (13) and the Oort cloud (14).  Gravitational
perturbations may occasionally change their parking orbits and bring them
into the inner solar system. Various ``anomalies'' in the solar planetary
system can be explained by collisions (15), capture or scattering of such
objects by the planets.  Passage of such objects near Earth could have
generated on it gigantic tidal waves, large volcanic eruptions and drastic
changes in global climate and sea level. Moreover, upon crossing the
asteroids and Kuiper belts these planetary mass objects could have
diverted large meteorites and astroids into a collision course with Earth
(16,17).  Thus visits of distant planets may be the common cause for the
various catastrophes that were jointly responsible for the major
biological mass extinctions.

\section*{Distant Planetary-Mass Objects}   

The recently discovered gigantic asteroids (11) in the outer solar system
between the Kuiper belt and the Oort cloud, various extrasolar
observations (12,18-19) and theoretical considerations (20-22) suggest that
many planetary-mass objects may be present in distant solar rings.
Probably, they have been formed together with the sun, since star
formation commonly involves formation of a thin planar disk of material
possessing too high an angular momentum to be drawn into the nascent star
and a much thicker outer ring of material extending out to several hundred
AU. Evidence for this material has been provided by infrared 
photometry of
young stars and also by direct imaging (18). Moreover, the recently
discovered Jupiter-mass planets orbiting very near solar-like stars (19)
much closer than where their formation was expected to take place, led to
the conclusion (20) that very many planetary-mass objects must be formed
at large radii where most of the disk mass resides, and migrate inward as
a result of tidal interactions with the planetary disk. Furthermore,
analysis of the properties of axisymmetric planetary nebula led to the
conclusion that many gas-giant planets are present around most main
sequence stars (22). Indeed, recent observations of the nearest planetary
nebula, the Helix Nebula, with the  Space Telescope (12) discovered 
$\sim 3500$ objects with a comet like shape (termed Cometary Knots) and a
typical mass of about $10^{-5}M_\odot$ (for comparison, $M_{Earth}\approx
3\times 10^{-6}M_\odot$ and $M_{Jupiter}\approx9.6\times 10^{-4}M_\odot$). 
in a distant ring around the central star.  It is not clear whether they
contain a solid body or uncollapsed gas. They are observed at distances
comparable to our own Oort cloud of comets but they seem to be distributed
in a planar ring rather than in a spherical cloud like the Oort cloud.  It
is possible that these planetary-mass objects and the recently discovered
gigantic asteroids (11) in the outer solar system between the Kuiper belt
and the Oort cloud are the high mass end of the vastly more numerous low
mass comets. The planetary-mass objects are more confined to the ecliptic
plane while the very light ones are scattered by gravitational collisions
into a spherical Oort cloud.  Gravitational interactions can change their
parking orbits into orbits which may bring them into the inner solar
system. 

\section*{Planetary Collisions} 
Collision of planet-mass objects with the known planets may explain
various ``anomalies'' in the solar planetary system (23):  A
collision of a Mars-size object with Earth could have formed our
moon (15) and tilt the spin plane of Earth ($23.5^0$) relative to the
ecliptic.  Such a collision could have formed Pluto's moon Charon, tilted
Pluto's orbital plane relative to the ecliptic plane by the observed
$17.1^0$ and change it into its observed high eccentric orbit, $e=0.253$.
The inclination of Mercury's orbit by $7.0^0$ and its high eccentricity,
$e=0.206$, could also have resulted from a near encounter with a visiting
planet. Collision, or near encounters, with visiting planets could have
tilted the spin plane of the other planets/moons/satellites relative to
their orbital plane (Mars: $25^0$, Jupiter: $3^0$, Saturn: $25^0$,
Neptune: $28^0$, Uranus: $98^0$ and Venus: $178^0$). 
 
\noindent
Capture by the giant planets of passing 
miniplanets/moons/asteroids, or of fragments resulting from a collision or
a tidal disruption can explain: (a) why there are moons/satellites, such
as Phobos and Deimos of Mars, with completely different density and
composition than that of their planets, (b) why there are moons with
unexplained retrograde orbits, such as Triton around Neptune, (c) why the
orbit planes of some moons, minimoons and asteroids are tilted relative to
the equatorial plane of their planets, in particular those with retrograde
orbits around the heavy planets Jupiter, Saturn and Neptune, and those
with prograde but very eccentric orbits, such as Nereid of Neptune which
has the most highly eccentric orbit of any known planet or satellite.  Out
of the 61 known moons, the six with the retrograde orbits (four around
Jupiter, one around Uranus and one around Neptune) and about the same
number of prograde moons have a large inclination ($\sim 20^0$) with
respect to the equatorial plane of the planet which they orbit.  This is
consistent with the fact that prograde and retrograde approaches are
equally probable, but, tidal interactions between the heavy planets and
their retrograde moons bring them closer and lead eventually to their
tidal capture or disruption into rings like those around Saturn. 

\section*{Recent Encounters}
Most of these could have happened in the early solar system when the
collision rate was much higher, as it is evident from dating of moon and
planetary craters (10). However, there are observations which suggest
that near encounters of planet-like objects with the Earth-Moon system
have occurred also more recently. For instance, fossil corals show that
the rate of decrease in the number of diurnal rings during annual cycles,
i.e., the rate of decrease in the number of days in a year, has changed
suddenly into a slower rate at the end Devonian (10). Since the
lengthening of the day is due to slowing of the rotation of Earth by the
well understood moon's tidal forces, it implies that at the end Devonian
the moon-Earth distance increased suddenly by a significant fraction. Such
an increase could have been induced by a tidal pull in a nearby passage of
a visiting planet/moon (a crash into Earth induces a discontinuous change
in the length of the day rather than a continuous one).  The tidal pull
increases the Moon-Earth separation if its projected trajectory in the
Moon-Earth orbiting plane does not fall between them.  The maximal radial
kick to the moon is when the trajectory of the visiting planet is
perpendicular to the Earth-Moon line on the far side of the moon.  For a
passing distance $d$ it is \begin{equation} V_r \approx {2GM_p\over
v}\left[ {1\over d}-{1\over d+D}\right], \end{equation} where
$D=380000~km~s^{-1}$ is the Moon-Earth distance and $v$ is the velocity of
the visiting planet relative to the Earth-Moon system (if
$V_p(\infty)\approx 0$ then $V_p(1~AU)\approx \sqrt{2}V_E$, and
$(\sqrt{2}-1)V_E\leq v\leq (\sqrt{2}+1)  V_E $ with $V_E\approx
30~km~s^{-1}$). For $d\gg D$, this tidal velocity is quite small. But, if
the visiting planet passes near the moon in a direct orbit at a distance
$d\leq D$, then $V_r\sim 2GM_p/vd$. Such a tidal velocity can increase
significantly the moon's distance from Earth and divert it into an
eccentric and inclined orbit (as observed). The eccentricity will be
damped by terrestrial tidal forces.  For $d<0.1(M_p/M_E)D$, a planet can
even eject the moon from its Earth-bound orbit. Nearby passage can also
slightly change the Earth orbit around the sun and its spin orientation.
Even slight changes in the orbit or in the spin orientation of Earth can
have dramatic effects on climate, sea level and glaciers.

\section*{Tidal Effects} 
Visiting planets and moons that pass near Earth can cause also gigantic
tidal waves and intense volcanism.  For simplicity, we limit our
discussion here to a single visit and ignore `` multiple visits'' before
escape to ``infinity'' or, capture/disruption by the sun. (which will be
discussed in detail elsewhere). Although exact calculations of surface
tidal effects are beyond the scope of this paper, 
an approximate estimate of the flexing of
Earth by a passing planet can be easily obtained by neglecting the
rotation of Earth and the speed of the passing planet (of mass $M_p$) and
by assuming a quasi hydrostatic equilibrium. By balancing the terrestrial
forces against the tidal force one obtains a surface displacement along
the line of sight to the planet which is given approximately by (24)
\begin{equation} 
h\approx {3\over 4}{M_p\over M_E}\left({R_E\over
d}\right)^3R_E. 
\end{equation} 
The maximal land tide due to the moon is $27~cm$.  However, a visiting
planet with a typical mass like that of the Cometary Knots (12), which
passes near Earth at a distance comparable to the Earth-Moon distance,
produces gigantic oceanic and continental tidal waves which are a few hundred
times higher than those induced by the moon.  Scaling of the tides
produced by the moon to those produced by visiting planets is justified
for approximate estimates since the duration of the strong tidal
acceleration by the moon, which is determined by the Earth rotation, is
similar to $t\sim d/v$, the passage time of a visiting planet at a passing
distance $d\sim D$ from Earth, with a velocity $(\sqrt{2}-1)V_E\leq v\leq
(\sqrt{2}+1)V_E$ relative to Earth.  Oceanic tidal waves, more than 1 km
high, can flood vast areas of continental land and devastate sea life and
land life near continental coasts. The spread of ocean waters by the giant
tidal wave over vast areas of land and near the polar caps will enhance
glaciation and sea regression.

Flexing Earth by $h\sim 100~m$ will deposit in it $\sim \alpha
GM_E^2h/R^2\approx 10^{34} erg$ , where $\alpha\sim 0.1$ is a geometrical
factor. It is approximately the heat release within Earth
during $10^6~y$ by radioactive decays. The flexing of Earth and the
release of such a large energy in a very short time upon contraction might
have triggered the gigantic volcanic eruptions that produced the Siberian
basalts flood at the time of the P/T extinction and the Deccan basalts
flood at the time of the K/T extinction (3,4). More distant
encounters and/or smaller visiting planets could have caused smaller
extinctions and sea regressions without massive volcanic eruptions. 

\section*{Planet Accretion}
A reliable estimate of the masses and the flux of the visiting
planets/planetesimals is not possible yet. However, it is tempting to
estimate them from the assumption that the unaccounted energy source of
Jupiter and its tilted spin plane relative to its orbital plane are both
due to accretion of visiting planets/moons.  An accretion rate of
$\dot{M}\approx 5.8\times 10^{-12}M_J~y^{-1}$, in addition to its
absorption of 55\% of the incident sun light, can explain its effective
$134K$ surface temperature. It implies that Jupiter has accreted $\approx
2.65\%$ of its mass during the $t_\odot\approx 4.57~Gy$ lifetime of the
solar system. Jupiter's effective capture cross section is,
$\sigma_J\approx \pi R_J^2(1+ M_JD_J/M_\odot R_J)$, where we assumed that
visiting planets arrive near Jupiter with a free fall velocity in the
sun's gravitational field, $V_p\approx \sqrt{2GM_\odot/D_J}$ with
$D_J=5.2AU$ being Jupiter's distance from the sun.  The impacts are from
random directions in the orbital plane. If the visiting planets have a
mean mass $M_p$ then the number of random impacts are $N_J\approx
0.0265M_J/M_p$.  Each random impact deposits on the average $\sim
(2/3)^{3/2}M_p V_J R_J$ of angular momentum perpendicular to Jupiter's
spin, $S\approx I_J\Omega_J\approx (2/5)M_J R_J^2\Omega_J$, where $I_J$
and $\Omega_J$ are its moment of inertia and angular rotational velocity,
respectively, and $V_J$ is the planet's impact velocity. Thus, the random
deposition of orbital angular momentum in Jupiter during $t_\odot$, could
have tilted its spin by a mean angle
\begin{equation}
sin\theta_{J}\approx
\sqrt{N_J}M_p{\left ({2\over 3}\right)^{3/2}
R_J\left[{2GM_J\over R_J}
+{2GM_\odot\over D_J}\right]^{1/2}\over I_J\Omega_J}.
\end{equation} 
From the observed $3.13^0$ tilt of Jupiter's spin and the accreted mass,
we infer that $M_p\approx 0.5M_E$ and $N_J\approx 16$. Similar estimates
for other planets, although yielding the correct order of magnitude
for the tilts of their spins, are
less reliable because the inferred number of accreted planets is too
small. 

\noindent
The sun is the main ``planet sweeper'' of the solar system. Its $7^0$ spin
tilt could have been produced by the impact of $\sim 3\times 10^4$ planets
of characteristic mass $\sim 0.5M_E$ (we ignore angular momentum loss by
the solar wind).  This means that the sun has accreted $\sim5\%$ of its
mass after its formation, at a rate of $\sim 7$ planets per $My$.  In each
capture episode, $\sim 6\times 10^{42}~erg$ of gravitation energy is
released in the sun's convective layer. It produces optical and X-ray
flashes at a rate $\sim 7~\times 10^{-6}~L_\odot^{-1}y^{-1}$ for sun-like
stars. It also causes a significant luminosity rise for an extended time
which may have induced climatic and sea level changes on Earth, and 
extinctions of species which could not have adapted to large environmental
changes. The predicted rate, is consistent with the observed rate of large
changes in $^{18}$O concentration in sea water sediments which record
large changes in sea water level (25) and in the total volume of glaciers.
Other implications of planet-sun and planet-star collisions 
will be discussed elsewhere.

\section*{Biological Mass Extinctions}
Using our inferred planet flux from Jupiter and its collimation by the
sun, we obtain that a ``visiting rate'' of once every $t_v=100~My$ for
planets with $M_p\sim 0.5M_E $ which fall towards the sun implies a
passing distance $d\leq (t_\odot/t_v)^{1/2}
[R_JM_JD_\odot/N_JM_\odot]^{1/2}\approx 170000~km$ from Earth. Then, from
eq. 2 we conclude that such visits produce land tides of $h\geq
125~m$ and ocean tidal waves which can reach $1km$ height.  The
combination of ocean tidal waves, volcanic eruptions, meteoritic impacts and
environmental and climatological changes can explain quite naturally the
biological and time patterns of mass extinctions. For instance, the giant
tidal waves devastate life in the upper oceans layers and on low lands
near coastal lines. They cover large land areas with sea water, spread
marine life to dry on land after water withdraw, and sweep land life into
the sea. They flood sweet water lakes and rivers with salt water and erode
the continental shores where most sea bed marine life is concentrated. 
Amphibians, birds and inland species, can probably survive the ocean tide.
This may explain their survival after the K/T extinction.  Survival at
high altitude inland sites may explain the survival of some inland
dinosaurs beyond the K/T border. Volcanic eruptions block sunlight,
deplete the ozone layer, and poison the atmosphere and the sea with acid
rain. Drastic sea level, climatic and environmental changes inflict
further delayed blows to marine and continental life.  But, high land life
in fresh water rivers which are fed by springs, that is not so sensitive
to temperature and climatic conditions has better chances to survive the
tidal waves, the volcano poisoning of sea water, and the drastic sea
level, environmental and climatic changes. 

\section*{conclusions} 
In spite of intensive studies it is still not known
what caused the biological mass extinctions and whether they were caused
by a single or a combination of extinction mechanisms. So far no single
mechanism has provided a satisfactory explanation of the complex
biological extinction patterns in sea and on land, the rate of mass
extinctions and their correlation with the largest volcanic eruptions, sea
regression and meteoritic impacts (1-5). These include
astrophysical extinction mechanisms such as meteoritic impacts (6)
passage of the solar system through molecular/dark matter
clouds (26), supernova explosions (27-29) gamma ray
bursts (30) and cosmic ray jets (31). However, visiting planets
offer a simple solution to the puzzling correlations between mass
extinctions, meteoritic impacts, volcanic eruptions, sea regression and
climatic changes as documented in geological records.  The hypothesis that
a substantial amount of mass is present in the outer solar system (and in
most other solar-like systems) in the form of planet-mass objects may be
tested in the near future by advanced gravitational lensing experiments
and by space based IR interferometric observations. Other consequences
for the solar system are currently under elaborate numerical and 
analytical investigations (32).

\end{document}